\documentclass{article}

\usepackage{arxiv}

\usepackage[utf8]{inputenc} 
\usepackage[T1]{fontenc}    
\usepackage{hyperref}       
\usepackage{booktabs}       
\usepackage{microtype}      
\usepackage{graphicx}
\graphicspath{{Images/}}
\usepackage[numbers,square]{natbib}
\usepackage{upgreek}
\usepackage{threeparttable}
\usepackage{subcaption}

\title{Thermophysical and mechanical properties of UFe$_2$ fabricated by spark plasma sintering}

\author{
Yifan Sun\thanks{Correspondence to: \texttt{sun.yifan.7r@kyoto-u.ac.jp}}\\
  Kyoto University, Japan \\
  \And
  Hironobu Nakamura \\
  Osaka University, Japan \\
  \And
  {Daisuke Okada}\\
  Osaka University, Japan\\
  \AND
  {Hiroaki Muta} \\
  Osaka University, Japan\\
  \And
  {Yuji Ohishi} \\
  Osaka University, Japan\\
  \And
  {Ken Kurosaki} \\
  Kyoto University, Japan\\
}

\renewcommand{\headeright}{}
\renewcommand{\undertitle}{}
\renewcommand{\shorttitle}{}

\begin{document}
\maketitle

\begin{abstract}
  Following the accident at the Fukushima Daiichi Nuclear Power Plant in 2011, core meltdown produced fuel debris whose safe retrieval and management require reliable thermophysical and mechanical property data. Among the metallic phases identified in the debris, the U--Fe system is particularly important because of the abundant iron originating from in-vessel stainless steel structures. However, within this system, the high-temperature thermophysical properties of UFe\textsubscript{2} have received relatively little attention, with most prior studies focusing on its magnetic and electronic properties. To fill this data gap in the literature, we fabricated dense, nearly single-phase polycrystalline UFe\textsubscript{2} by arc melting followed by spark plasma sintering, and characterized its thermal and mechanical properties from room temperature to 1073~K. Results show that the thermal conductivity of UFe\textsubscript{2} increased monotonically from 10~W~m$^{-1}$~K$^{-1}$ at 306~K to 25~W~m$^{-1}$~K$^{-1}$ at 1073~K, surpassing those of the iron intermetallics Fe\textsubscript{2}Zr and Fe\textsubscript{2}B at high temperatures. In addition, UFe\textsubscript{2} is mechanically more compliant, displaying a Young's modulus $E$ of 69~GPa, a shear modulus $G$ of 24~GPa, and a Vickers hardness $H_{\mathrm{V}}$ of 5.6~GPa, all well below those of both Fe intermetallics. Consequently, during decommissioning, thermal-management and structural evaluations should take into account the comparatively high-conductivity and mechanically compliant nature of UFe\textsubscript{2} within the heterogeneous fuel debris.
\end{abstract}

\keywords{Intermetallic \and U--Fe \and Fuel debris \and Thermal conductivity \and Mechanical properties}

\section{Introduction}

In 2011, the earthquake and tsunami that struck the Fukushima Daiichi Nuclear Power Plant (1F) caused a station blackout and the loss of reactor cooling. Core meltdown subsequently occurred, leaving solidified fuel debris at the bottom of the primary containment vessels. In the ongoing decommissioning process, accurate knowledge of the phase states as well as the thermal and mechanical properties of fuel debris is essential for its safe and smooth retrieval and storage management~\cite{nakayoshi2020review,kurata2022step,zubekhina2023key}. In particular, thermal properties such as thermal conductivity are fundamental for evaluating heat transfer within fuel debris and underpin the development of retrieval technologies---including core-boring and thermal-cutting tools~\cite{jaea2014fueldebris}.

Previous studies using simulated fuel debris and thermodynamic assessments have suggested that oxide, boride, and metallic phases may be formed in fuel debris~\cite{takano2014characterization}. For example, as representative oxide phases, (U,Zr)O\textsubscript{2} has been reported~\cite{takano2014characterization,takano2013high}; boride phases such as ZrB\textsubscript{2} and (Fe,Cr,Ni)\textsubscript{2}B, formed through reactions between control rod--derived B\textsubscript{4}C and in-vessel structural materials, have also been identified~\cite{takano2014characterization,steinbruck2010degradation,nagase1997b4c,nagase1997chemical,hofmann1990reaction}. For these phases, experimental studies on thermal conductivity and mechanical properties have been increasingly accumulated~\cite{fink1981thermophysical,song2023thermophysical,nakamori2016mechanical,nakamori2015mechanical}.

With respect to metallic phases, thermodynamic calculations of in-vessel fuel debris by Ikeuchi et al.\ indicate that, under the reducing conditions with an abundance of Fe, a fraction of the uranium can be reduced by zirconium and taken up into an Fe--Zr intermetallic, forming the Laves phase Fe\textsubscript{2}(Zr,U)~\cite{ikeuchi2020chemical,ikeuchi2013suggestion}. Such an Fe\textsubscript{2}Zr intermetallic has been observed experimentally in high-temperature reaction experiments between Fe--Zr melts and stainless steel~\cite{itoh2025experimental}, and its properties have been characterized~\cite{okada2019thermal}. On the other hand, the uranium end-member of this Laves phase, UFe\textsubscript{2}, has been studied mainly for its magnetic and electronic properties~\cite{zentko1979ferromagnetism,andreev2002spin,moseley2025expanding}, with little attention from the perspective of decommissioning.

While some experimental data are available for UFe\textsubscript{2}, its fundamental thermophysical properties remain only partially characterized. Apart from the high-temperature heat capacity reported by Rai and Raju~\cite{rai2013high} and the room-temperature elastic properties reported by Yamanaka et al.~\cite{yamanaka1998mechanical}, key properties such as the thermal expansion coefficient and thermal conductivity, crucial for decommissioning activities, have yet to be experimentally measured at elevated temperatures. To address this data gap, we synthesized dense, nearly single-phase UFe\textsubscript{2} by arc melting followed by spark plasma sintering (SPS), and characterized its thermal expansion coefficient, thermal conductivity, and various mechanical properties. The thermophysical property data of UFe\textsubscript{2} presented in this work are expected to serve as fundamental data to evaluate the thermal behavior of fuel debris and to contribute to the development of future decommissioning technologies. By comparing the obtained results with reported data for oxide and boride phases expected to be present in fuel debris, the role of the UFe\textsubscript{2} metallic phase in fuel debris is further discussed.

\section{Experimental methods}
\subsection{Sample synthesis and characterization}
UFe\textsubscript{2} samples were prepared by arc melting stoichiometric amounts of U (Nippon Nuclear Fuel Industry Co., Ltd.) and Fe (Furuuchi Chemical Co., Ltd.) under an Ar atmosphere at a pressure of 0.04~MPa. To improve homogeneity, each ingot was inverted and remelted four times. The ingot was then crushed for 10~min using a tungsten carbide mortar and pestle inside an Ar-filled glovebox to prevent oxidation, and the resulting powder was loaded into the SPS die and punch within the same glovebox. The powder was consolidated by SPS (SPS-515A, Sumitomo Coal Mining Co., Ltd.) to obtain dense bulk specimens. The sintering temperature was raised to 1273~K under an Ar flow of 200~mL~min$^{-1}$ and held for 10~min under a uniaxial pressure of 100~MPa.

Phase identification of the bulk UFe\textsubscript{2} samples was carried out by X-ray diffraction (XRD, Rigaku Ultima IV) using Cu~K$\alpha$ radiation over a $2\uptheta$ range of 20--120$^\circ$. High-temperature phase evolution was further examined by high-temperature X-ray diffraction (HT-XRD), in which powder samples were measured under flowing He from room temperature to 473~K at 25~K intervals. The lattice parameter at each temperature was determined from the HT-XRD patterns using the Cohen method, which corrects for the systematic peak shift arising from specimen-surface displacement, together with an instrumental zero offset correction based on the peak positions of a silicon standard (NIST SRM 640c); the room-temperature lattice parameter was taken from the 298~K measurement of this series. The room-temperature bulk density was calculated from the specimen mass and geometric dimensions, and the relative density was evaluated against the theoretical density derived from the room-temperature lattice parameter. From the temperature dependence of the HT-XRD lattice parameters, the linear thermal expansion coefficient was then obtained using the following equation:

\begin{equation}\label{eq:linear-thermal-expansion-coefficient}
\alpha_l = \frac{1}{l_0}\frac{l - l_0}{T - T_0}
\end{equation}

Here, the reference temperature $T_0$ is 298~K, and $l_0$ and $l$ are the lattice parameters at $T_0$ and at a given temperature $T$, respectively. Finally, surface morphology was observed using scanning electron microscopy (SEM, JSM-6500F, JEOL), and elemental distribution was analyzed by energy-dispersive X-ray spectroscopy (EDS, EX-23000BU, JEOL).

\subsection{Property measurements}
To evaluate the high-temperature phase stability and chemical reactivity of UFe\textsubscript{2}, thermogravimetric and differential thermal analysis (TG-DTA) was conducted. The measurements were performed using a TG-DTA2000SA system (BRUKER AXS) from room temperature to 1073~K under both air and flowing Ar atmospheres. Thermal diffusivity ($\alpha$) was measured by the laser flash method (LFA-457, NETZSCH) under an Ar flow of 200~mL~min$^{-1}$ from room temperature to 1073~K. At each temperature, three laser shots were performed and the average value was used. Thermal conductivity ($\kappa$) was calculated from $\kappa = \alpha C_{\mathrm{p}}\rho$, where $\rho$ is the measured density corrected for thermal expansion determined via HT-XRD, assuming it holds over the entire measured range to 1073~K, and $C_{\mathrm{p}}$ is the specific heat, obtained by refitting the heat-capacity curve reported by Rai and Raju~\cite{rai2013high}. The resulting thermal conductivity was corrected to the fully dense value using the Maxwell--Eucken relation~\cite{morimoto2008thermal},
\begin{equation}\label{eq:maxwell-eucken}
\kappa_{\mathrm{p}} = \kappa_{0}\,\frac{1-P}{1+\beta P},
\end{equation}
where $\kappa_{\mathrm{p}}$ and $\kappa_{0}$ are the thermal conductivities of the porous and fully dense material, respectively, $P$ is the porosity, and $\beta = 0.5$ is the correction coefficient. 

The longitudinal ($v_{L}$) and transverse ($v_{S}$) sound velocities of UFe\textsubscript{2} were measured using the ultrasonic pulse-echo method (Echometer 1062, Nihon Matech Co., Ltd.). The shear modulus ($G$), Young’s modulus ($E$), bulk modulus ($B$), Poisson’s ratio ($\nu$), and Debye temperature ($\theta_{D}$) were calculated using the following equations:

\begin{equation}\label{eq:rigidity}
G = \rho v_S^2 
\end{equation}
\begin{equation}\label{eq:young}
E = G \frac{3v_L^2 - 4v_S^2}{v_L^2 - v_S^2} 
\end{equation}
\begin{equation}\label{eq:bulk}
B = \rho \left( v_L^2 - \frac{4}{3} v_S^2 \right) 
\end{equation}
\begin{equation}\label{eq:poisson}
\nu = \frac{1}{2} \frac{v_L^2 - 2v_S^2}{v_L^2 - v_S^2} 
\end{equation}
\begin{equation}\label{eq:debye}
\theta_D = \frac{h}{k_B}
\left[
\frac{9\rho n N_A}{4\pi M \left( 2v_S^{-3} + v_L^{-3} \right)}
\right]^{1/3}
\end{equation}

where $\rho$ is the density, $h$ is Planck’s constant, $k_B$ is Boltzmann’s constant, $n$ is the number of atoms per formula unit (3 for UFe\textsubscript{2}), $N_A$ is Avogadro’s number, and $M$ is the molar mass.

Vickers hardness was measured at room temperature (298~K) using a microhardness tester (HMV-G305, Shimadzu) under a load of 9.807~N held for a dwell time of 10~s, with the reported value averaged over ten indentations. The Vickers hardness $H_{\mathrm{V}}$ (kgf~mm$^{-2}$) was obtained from $H_{\mathrm{V}} = 0.1891\,F/d^{2}$, where $F$ is the applied load (N) and $d$ is the indentation diagonal length (mm), and the values were subsequently converted to SI units (GPa).

\section{Results}
\subsection{Sample characterization}
\begin{figure}[htbp]
  \centering
  \begin{subfigure}[t]{0.8\textwidth}
    \centering
    \includegraphics[width=\linewidth]{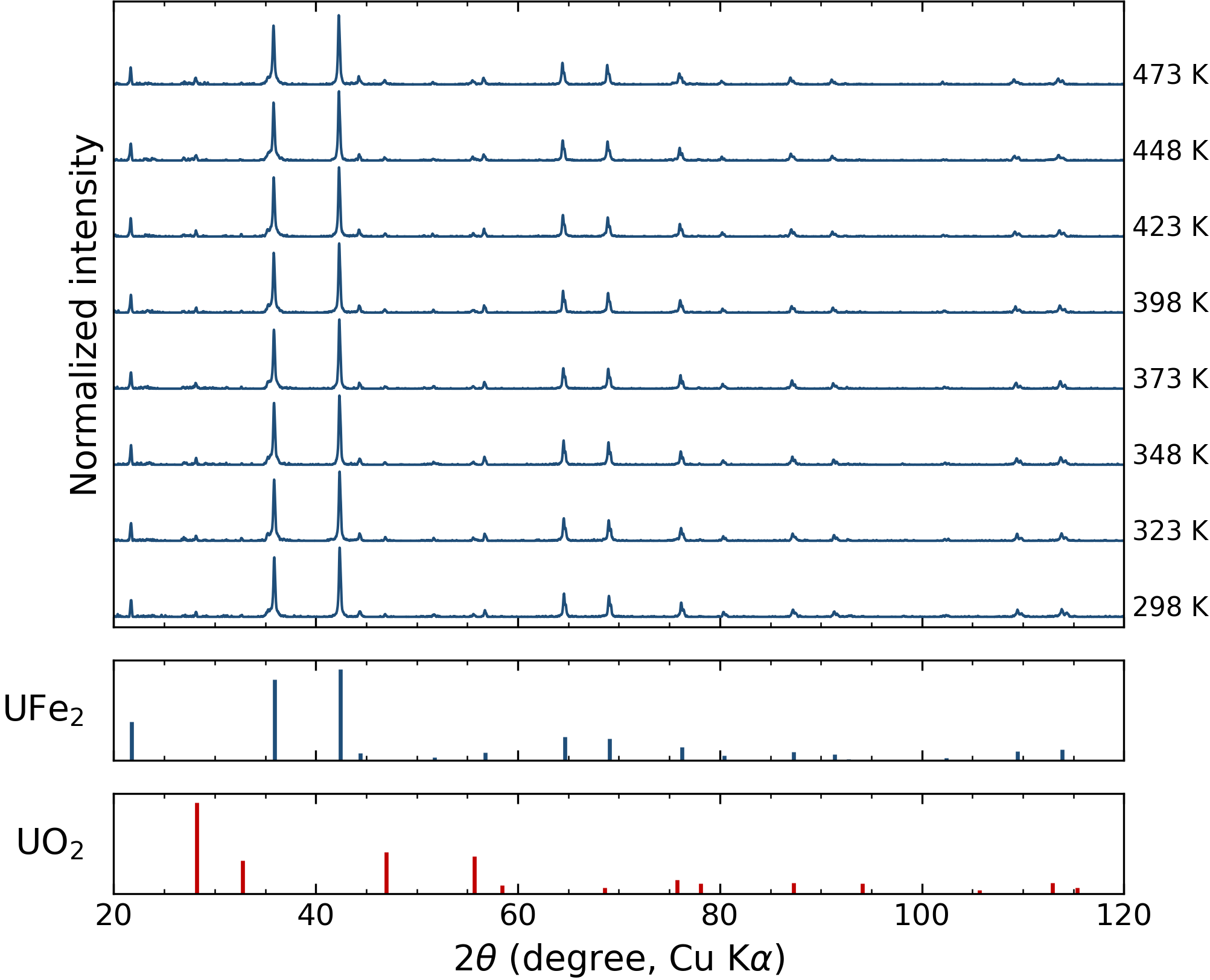}
    \caption{}
    \label{fig:ht-xrd}
  \end{subfigure}
  \hfill
  \begin{subfigure}[t]{0.6\textwidth}
    \centering
    \includegraphics[width=\linewidth]{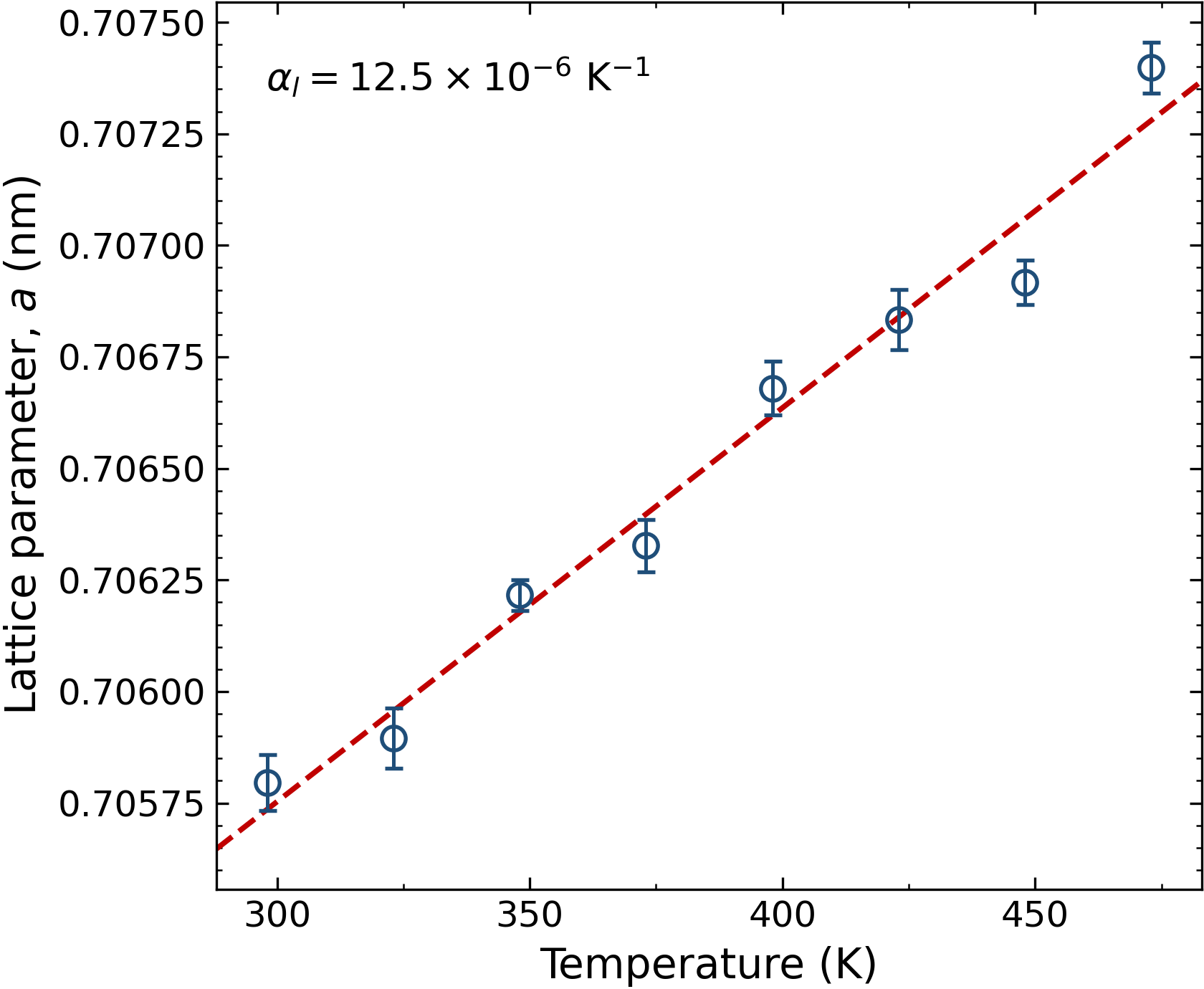}
    \caption{}
    \label{fig:thermal-expansion}
  \end{subfigure}
  \caption{\textbf{XRD patterns of the UFe\textsubscript{2} powdered sample.} (a) HT-XRD patterns from room temperature to 473~K; the UFe\textsubscript{2} peaks shift systematically toward lower angles owing to lattice expansion, with no new impurity phases forming, and the reference patterns for UFe\textsubscript{2}~\cite{baenziger1950compounds} and UO\textsubscript{2}~\cite{desgranges2009neutron} are shown for comparison. (b) Temperature dependence of the lattice parameter $a(T)$, with a linear fit (dashed) giving $\alpha_l = 12.5\times10^{-6}~\mathrm{K^{-1}}$.}
  \label{fig:htxrd}
\end{figure}

\begin{figure}[htbp]
  \centering
  \includegraphics[width=0.95\linewidth]{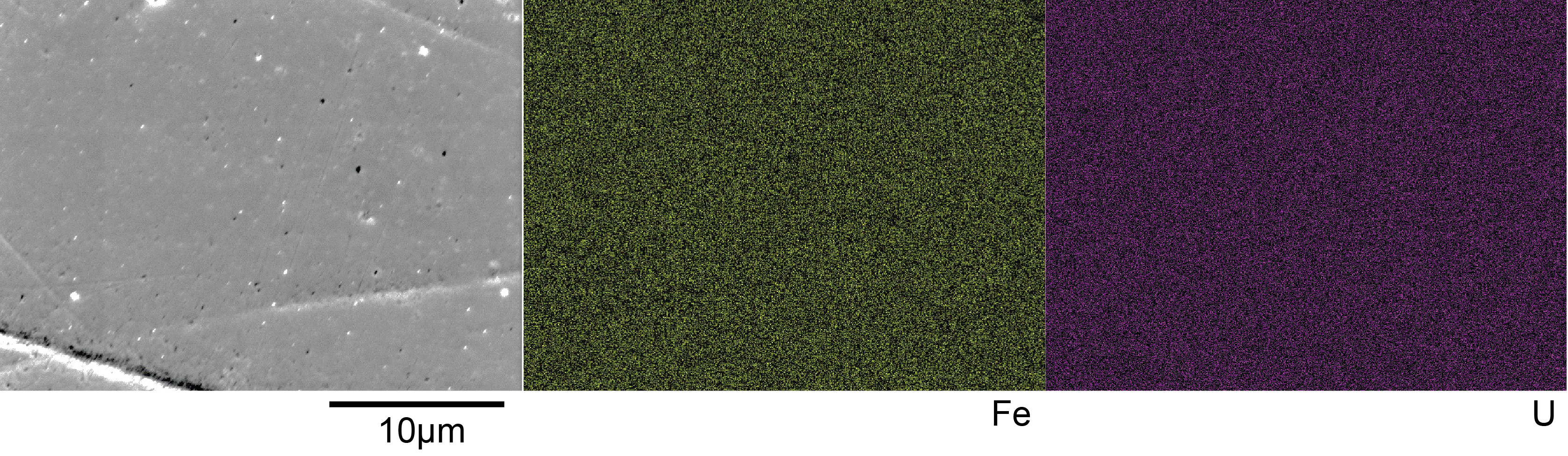}
  \caption{\textbf{SEM-EDS images of the prepared UFe\textsubscript{2} sample after SPS.}}
  \label{fig:sem-eds}
\end{figure}

Figure~\ref{fig:ht-xrd} shows the powder XRD pattern of the UFe\textsubscript{2} sample after SPS. At room temperature, most diffraction peaks can be indexed to the cubic UFe\textsubscript{2} phase~\cite{baenziger1950compounds}, confirming that the sample is predominantly single-phase UFe\textsubscript{2}, while a few weak additional reflections are attributed to a minor UO\textsubscript{2} secondary phase, consistent with the UO\textsubscript{2} reference pattern~\cite{desgranges2009neutron}. This composition is consistent with the SEM-EDS maps in Figure~\ref{fig:sem-eds}, which reveal a homogeneous distribution of U and Fe throughout the sample. The room-temperature lattice parameter of the cubic UFe\textsubscript{2} phase is calculated to be 0.7058~nm, in good agreement with previously reported values~\cite{baenziger1950compounds} (Table~\ref{tab:UFe2_lattice_density}). The sample also achieved a high relative density of 95\%, consistent with the dense microstructure seen in the SEM image (Figure~\ref{fig:sem-eds}). These results collectively confirm that a dense, nearly single-phase UFe\textsubscript{2} sample was successfully fabricated for the subsequent property characterization.

\begin{table}[htbp]
  \centering
  \caption{Lattice parameter and density of the synthesized UFe\textsubscript{2} sample}
  \label{tab:UFe2_lattice_density}
  \small
  \begin{tabular}{lccc}
    \toprule
    &
    \begin{tabular}{c}
      Lattice parameter\\
      (nm)
    \end{tabular}
    &
    \begin{tabular}{c}
      Measured density\\
      (g~cm$^{-3}$)
    \end{tabular}
    &
    \begin{tabular}{c}
      Relative density\\
      (\%)
    \end{tabular}
    \\
    \midrule
    This work & 0.7058 & 12.59 & 95 \\
    Reference~\cite{baenziger1950compounds} & 0.7059 & --- & --- \\
    \bottomrule
  \end{tabular}
\end{table}

The HT-XRD patterns, measured under flowing He from room temperature to 473~K (Figure~\ref{fig:ht-xrd}), showed no new phases forming over this temperature range, and the minor UO\textsubscript{2} peaks already present in the SPS sample remained relatively stable. With increasing temperature, the UFe\textsubscript{2} peaks shifted continuously toward lower angles, as expected from thermal expansion of the lattice. The temperature dependence of the lattice parameter from HT-XRD is shown in Figure~\ref{fig:thermal-expansion}. A linear fit of $a(T)$ (Equation~\ref{eq:linear-thermal-expansion-coefficient}) gives a linear thermal expansion coefficient $\alpha_l = 12.5\times10^{-6}~\mathrm{K^{-1}}$ over 298--473~K. For the cubic structure this translates into a volumetric coefficient $\alpha_V = 3\alpha_l = 37.5\times10^{-6}~\mathrm{K^{-1}}$.

\subsection{High-temperature phase stability}

\begin{figure}[htbp]
  \centering
  \begin{subfigure}[t]{0.49\textwidth}
    \centering
    \includegraphics[width=\linewidth]{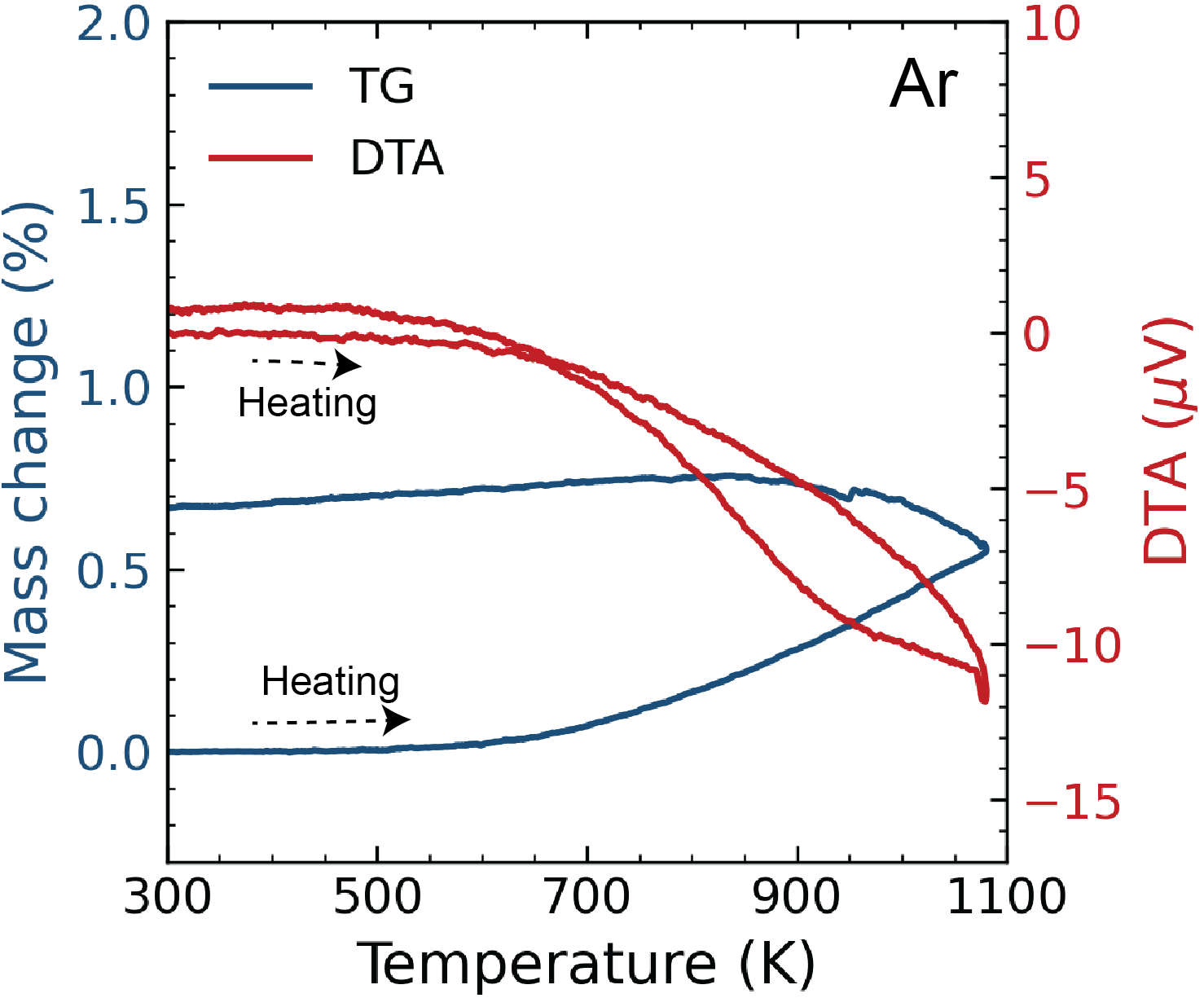}
    \caption{}
    \label{fig:tg-dta-ar}
  \end{subfigure}
  \hfill
  \begin{subfigure}[t]{0.49\textwidth}
    \centering
    \includegraphics[width=\linewidth]{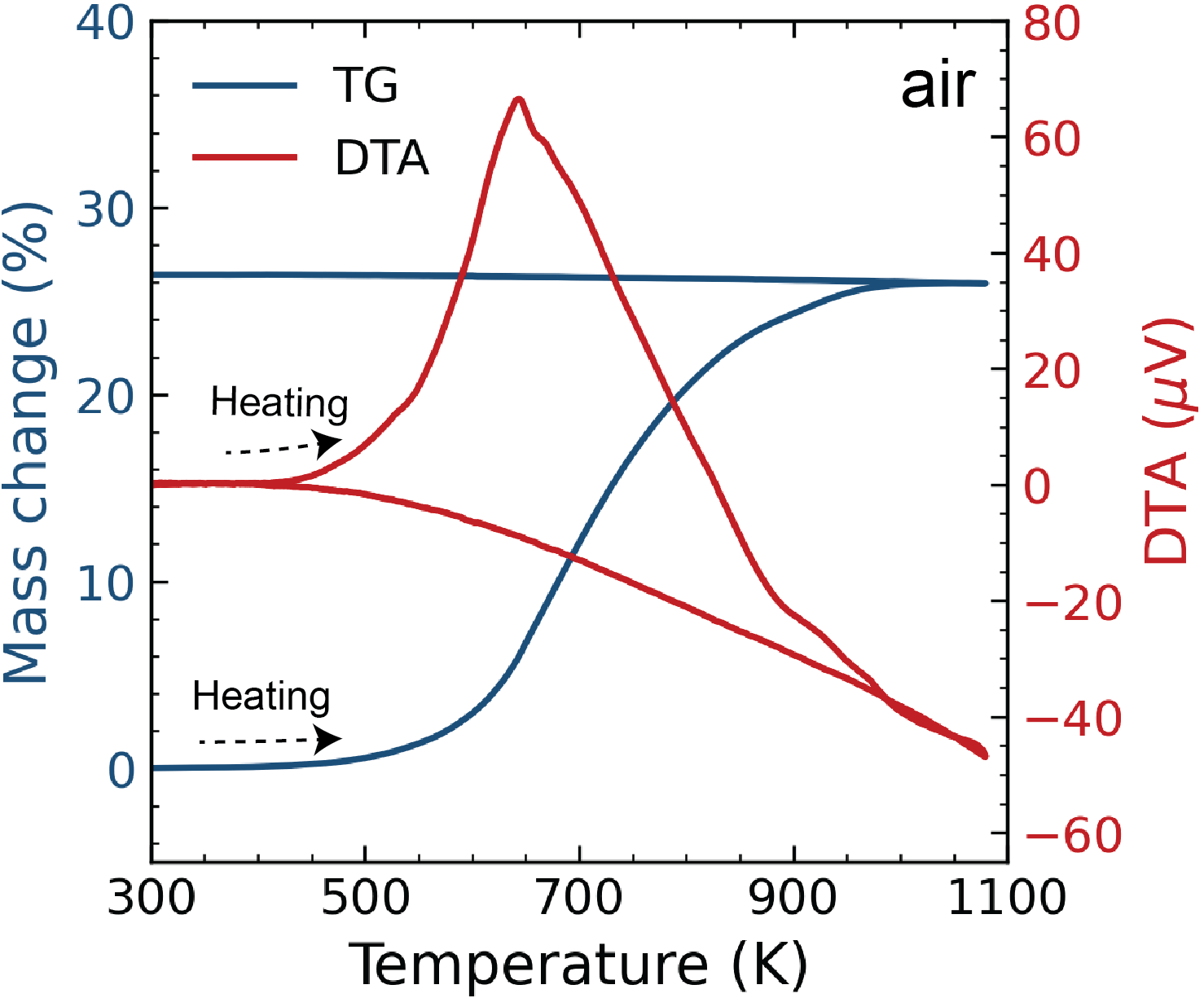}
    \caption{}
    \label{fig:tg-dta-air}
  \end{subfigure}
  \caption{\textbf{TG-DTA results of the UFe\textsubscript{2} samples, with arrows indicating the heating direction.} (a) Under flowing Ar, the mass change is negligible and no exothermic or endothermic peaks are observed up to $\sim$1073~K. (b) In air, oxidation produces a strong exothermic peak ($\sim$640~K) and an irreversible mass gain.}
  \label{fig:tg-dta}
\end{figure}

To further investigate the high-temperature stability of UFe\textsubscript{2} above 473~K, in both air and an inert atmosphere, we performed TG-DTA analysis, shown in Figure~\ref{fig:tg-dta}. Under flowing Ar (Figure~\ref{fig:tg-dta-ar}), no endothermic or exothermic peaks were detected up to $\sim$1073~K, aside from a gradual DTA baseline drift. This indicates that the fabricated UFe\textsubscript{2} sample undergoes no phase transformation or decomposition over this range, consistent with the U--Fe phase diagram~\cite{leibowitz1991thermodynamics} as well as our HT-XRD data. The sample's mass change was negligible ($\sim$1\%), indicating only slight oxidation by residual oxygen during the measurement.

In contrast, the high-temperature behavior of UFe\textsubscript{2} is markedly different in air (Figure~\ref{fig:tg-dta-air}). Analysis of the first and second derivatives of the TG and DTA curves shows that both signals began to deviate upward only slightly above $\sim$450~K. The exothermic DTA feature became more pronounced above $\sim$500~K and developed into a sharp peak at $\sim$640~K. The accompanying mass gain was initially small, remaining below 1\% up to $\sim$530~K, then picked up pace at around 550~K and reached its maximum rate near 680~K. This concurrent mass gain identifies the exotherm as the oxidation of UFe\textsubscript{2}, with the sample mass increasing by approximately 26\%. The uptake leveled off near 1000~K, indicating that oxidation was essentially complete by that temperature, and the mass remained roughly constant during cooling. The magnitude of this mass increase strongly suggests that the UFe\textsubscript{2} sample transformed into a mixture of U\textsubscript{3}O\textsubscript{8} and Fe\textsubscript{2}O\textsubscript{3}, based on their theoretical stoichiometric balances.

\subsection{Thermal conductivity}\label{subsec:thermal-conductivity}

\begin{figure}[htbp]
  \centering
  \includegraphics[width=0.6\linewidth]{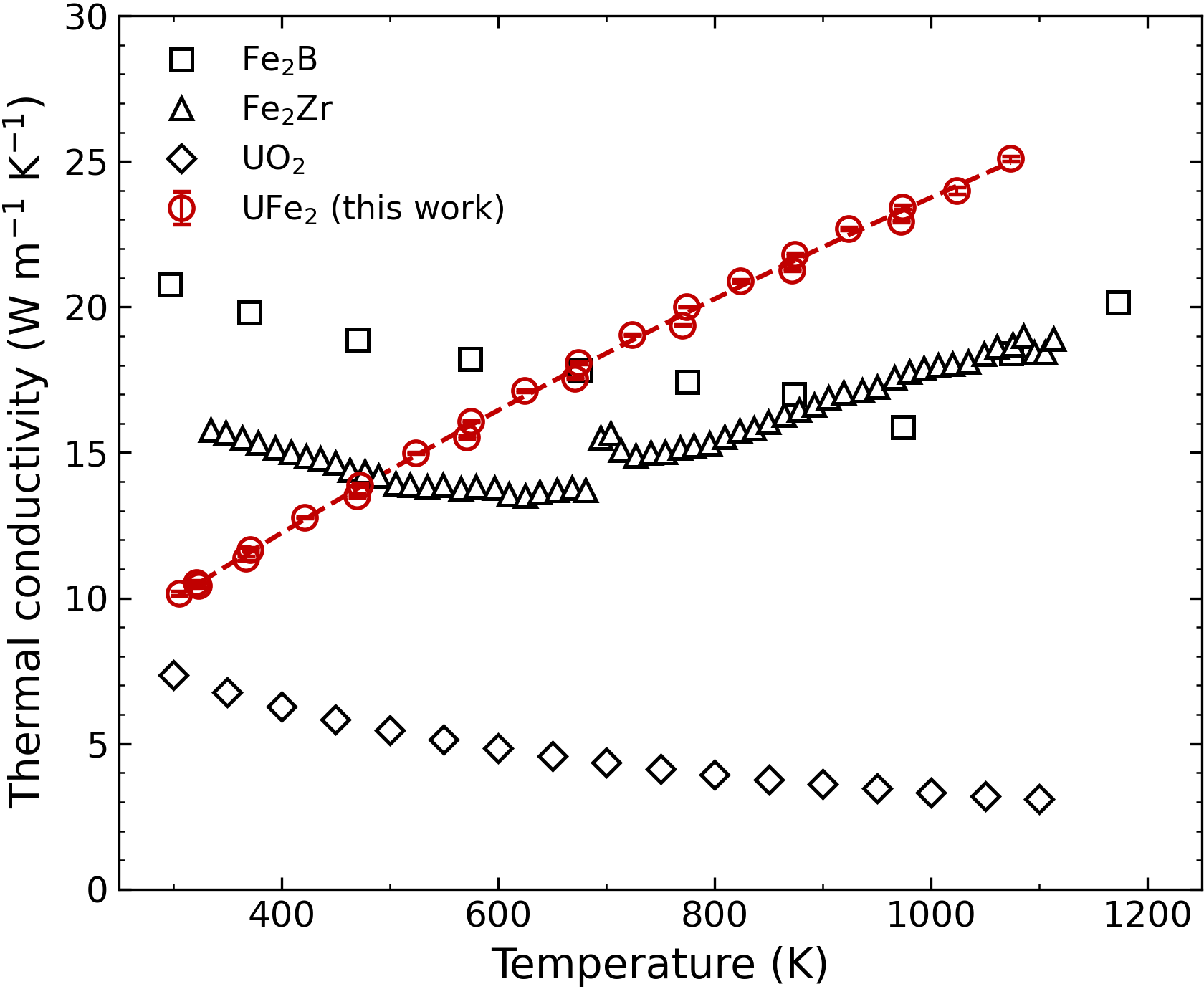}
  \caption{\textbf{Temperature dependence of the thermal conductivity of UFe\textsubscript{2}.} Literature values for UO\textsubscript{2}~\cite{ronchi1999thermal}, Fe\textsubscript{2}B~\cite{nakamori2016mechanical}, and Fe\textsubscript{2}Zr~\cite{okada2019thermal} are shown for comparison. The Fe\textsubscript{2}B and Fe\textsubscript{2}Zr values were extracted from the scatter plots in the respective references.}
  \label{fig:thermal-conductivity}
\end{figure}

The calculated thermal conductivity of UFe\textsubscript{2} is shown in Figure~\ref{fig:thermal-conductivity} together with the literature values for UO\textsubscript{2}~\cite{ronchi1999thermal} and the Fe intermetallics Fe\textsubscript{2}B~\cite{nakamori2016mechanical} and Fe\textsubscript{2}Zr~\cite{okada2019thermal}. Here, the heat capacity was obtained by refitting the experimental data of Rai and Raju~\cite{rai2013high}, as their proposed equation does not reproduce their reported heat capacity curve. The thermal conductivity of UFe\textsubscript{2} did not exhibit hysteresis between the heating and cooling runs, showing a moderate thermal conductivity of 10~W~m$^{-1}$~K$^{-1}$ at 306~K, which increased monotonically with temperature and reached 25~W~m$^{-1}$~K$^{-1}$ at 1073~K. The temperature dependence of the thermal conductivity can be well described by the empirical relation that has been applied to related uranium compounds, including UCo~\cite{sun2026uco} and UB\textsubscript{4}~\cite{kardoulaki2020thermophysical},

\begin{equation}\label{eq:kappa-fit}
\kappa(T) = \frac{T}{cT + d} + e,
\end{equation}

where the first term represents the electronic contribution and $e$ is a small constant offset accounting for the residual contribution. The fit (dashed line in Figure~\ref{fig:thermal-conductivity}) reproduces the measured data well ($R^2 = 0.998$), with fitted constants $c = 1.03\times10^{-2}$, $d = 36.5$, and $e = 2.38$ ($T$ in K, $\kappa$ in W~m$^{-1}$~K$^{-1}$).

As with other Fe intermetallics, the thermal conductivity of UFe\textsubscript{2} is consistently higher than that of UO\textsubscript{2}, and markedly so at elevated temperatures owing to the electronic contribution to heat transport. However, unlike Fe\textsubscript{2}Zr and Fe\textsubscript{2}B, whose thermal conductivities remain relatively stable at around 15--20~W~m$^{-1}$~K$^{-1}$ (showing a shallow minimum before rising), that of UFe\textsubscript{2} increases steadily with temperature, resembling the behavior reported for UB\textsubscript{4}~\cite{kardoulaki2020thermophysical}. We attribute this difference partly to the inherently lower lattice thermal conductivity in UFe\textsubscript{2}, arising from its much heavier constituent atoms, which lower the phonon velocity. The resulting lower Debye temperature ($\approx$190~K, shown below) lies far below the measured temperature range, where the lattice contribution has already fallen off with temperature, leaving the rising electronic contribution to dominate. Overall, UFe\textsubscript{2} exhibits a thermal conductivity comparable to, but somewhat higher than, those of the other Fe intermetallics expected in fuel debris, particularly at elevated temperatures. These results demonstrate that even among the Fe intermetallics that may coexist in fuel debris, the thermal conductivities and their temperature dependences can differ considerably, providing important insights for evaluating thermal behavior under high-temperature conditions.

\subsection{Elastic and mechanical properties}

The longitudinal and transverse sound velocities of the UFe\textsubscript{2} specimen were measured to be 3709~m~s$^{-1}$ and 1405~m~s$^{-1}$, respectively. The resulting mechanical properties, calculated using Equations~\ref{eq:rigidity}--\ref{eq:debye} together with the Vickers hardness, are summarized in Table~\ref{tab:UFe2_mechanical-properties}, alongside literature values for UFe\textsubscript{2}~\cite{yamanaka1998mechanical} and for the other phases relevant to decommissioning, such as Fe\textsubscript{2}Zr~\cite{okada2019thermal}, Fe\textsubscript{2}B~\cite{nakamori2016mechanical}, and UO\textsubscript{2}~\cite{kato2015thermal,yamada1998mechanical}.

\begin{table}[h]
  \centering
  \begin{threeparttable}
  \caption{Young’s modulus ($E$), bulk modulus ($B$), shear modulus ($G$), Poisson’s ratio ($\nu$), Debye temperature ($\theta_D$), and Vickers hardness ($H_{\mathrm{V}}$) of the sintered UFe\textsubscript{2} sample, along with literature values for UFe\textsubscript{2}~\cite{yamanaka1998mechanical}, Fe\textsubscript{2}Zr~\cite{okada2019thermal}, Fe\textsubscript{2}B~\cite{nakamori2016mechanical}, and UO\textsubscript{2}~\cite{kato2015thermal,yamada1998mechanical}.}
  \label{tab:UFe2_mechanical-properties}
  \small
  \begin{tabular}{lccccc}
    \toprule
    &
    \begin{tabular}{c}
      UFe\textsubscript{2}\\
      (This work)
    \end{tabular}
    &
    \begin{tabular}{c}
      UFe\textsubscript{2}\\
      \cite{yamanaka1998mechanical}
    \end{tabular}
    &
    \begin{tabular}{c}
      Fe\textsubscript{2}Zr\\
      \cite{okada2019thermal}
    \end{tabular}
    &
    \begin{tabular}{c}
      Fe\textsubscript{2}B\\
      \cite{nakamori2016mechanical}
    \end{tabular}
    &
    \begin{tabular}{c}
      UO\textsubscript{2}\\
      \cite{kato2015thermal,yamada1998mechanical}
    \end{tabular}
    \\
    \midrule
    $E$ (GPa) & 69 & 62.4 & 186 & 252 & 219.3 \\
    $B$ (GPa) & 138 & 134 & 129 & 182 & 202.1 \\
    $G$ (GPa) & 24 & 21.9 & 73.9 & 107 & 83.1 \\
    $\nu$ & 0.42 & 0.42 & 0.26\tnote{a} & 0.179 & 0.32 \\
    $\theta_D$ (K) & 190 & 178 & 432 & 620 & 384 \\
    $H_{\mathrm{V}}$ (GPa) & 5.63 $\pm$ 0.14 & 0.78 & 8.3 $\pm$ 0.1 & 12 & 4.6 \\
    \bottomrule
  \end{tabular}
  \begin{tablenotes}
    \footnotesize
    \item[a] Derived from the reported $E$ and $G$; $\nu$ was not tabulated in the reference.
  \end{tablenotes}
  \end{threeparttable}
\end{table}

Overall, our elastic moduli agree well with those reported by Yamanaka et al.~\cite{yamanaka1998mechanical}, the differences falling within experimental uncertainty. The one notable exception is the Vickers hardness, for which our value is roughly seven times higher than theirs. The discrepancy with the present study is surprising considering that Yamanaka et al.\ employed a smaller indentation load (4.905~N), which, through the indentation size effect, would tend to raise rather than lower the measured hardness. Compared with related Fe-based compounds, such as FeB ($\approx$12.5~GPa)~\cite{ohishi2019thermophysical}, Fe\textsubscript{2}Zr~\cite{okada2019thermal}, and Fe\textsubscript{2}B~\cite{nakamori2016mechanical}, their reported value of 0.78~GPa for UFe\textsubscript{2} is notably low for an intermetallic. Consequently, although the origin of this discrepancy cannot be established with certainty, we consider the present value of $5.63 \pm 0.14$~GPa to be the more reliable one, as it lies within the range of $\sim$5--12~GPa spanned by the other uranium intermetallic compounds~\cite{yamanaka1998mechanical} and Fe-based compounds.

Compared with the other phases expected in fuel debris, UFe\textsubscript{2} offers markedly less resistance to deformation: its Young's modulus (69~GPa) and shear modulus (24~GPa) are several times lower than those of UO\textsubscript{2} and Fe\textsubscript{2}B, and its Vickers hardness (5.63~GPa) is roughly half that of Fe\textsubscript{2}B. Its high Poisson's ratio (0.42), in turn, indicates a far more ductile mechanical response. The low Young's modulus reflects weaker interatomic bonding in UFe\textsubscript{2} which, together with its heavier constituent atoms, leads to the markedly low Debye temperature of 190~K. These comparisons underline that the mechanical properties of the phases that may coexist in fuel debris can differ substantially. UFe\textsubscript{2} is notably compliant and ductile, not only relative to the stiff, brittle oxide and boride phases, but also among the Fe-based intermetallics. This heterogeneity in mechanical behavior should therefore be taken into account when assessing the structural integrity of fuel debris.

\section{Discussion}

\subsection{Thermal transport behavior}

Based on the experimentally determined thermal expansion and sound velocity data, the Grüneisen parameter of UFe\textsubscript{2} can be evaluated directly, which in turn allows the lattice and electronic contributions to the thermal conductivity to be examined. Expressed in terms of the measured sound velocities, the thermodynamic Grüneisen parameter is given by
\begin{equation}\label{eq:gruneisen}
\gamma = \frac{\alpha_V M \left(v_L^2 - \frac{4}{3}v_S^2\right)}{C_{\mathrm{p}}},
\end{equation}
where $\alpha_V$ is the volumetric thermal expansion coefficient, $M$ the molar mass, and $C_{\mathrm{p}}$ the molar heat capacity. Using $\alpha_V = 37.5\times10^{-6}~\mathrm{K^{-1}}$, $v_L = 3709$~m~s$^{-1}$, $v_S = 1405$~m~s$^{-1}$, $M = 349.72$~g~mol$^{-1}$, and the refitted heat capacity of Rai and Raju~\cite{rai2013high}, we obtain $\gamma = 1.85$ at 300~K. Our value is somewhat smaller than the $2.25$ adopted by Rai and Raju in their quasi-harmonic Debye--Grüneisen modeling~\cite{rai2013high}. The difference stems primarily from how the underlying bulk modulus is obtained. Our adiabatic value, measured directly from sound velocities ($B_{S} = 137.6$~GPa), agrees closely with the earlier ultrasonic measurement of Yamanaka et al.~\cite{yamanaka1998mechanical}. For Rai and Raju, however, their estimates derive from re-analyzing the high-pressure diamond-anvil data of Itié et al.~\cite{itie1986}. They obtained $2.13$ from the thermodynamic relation using $B_{T} = 201$~GPa, and their adopted value $2.25$ from Slater's empirical formula using its pressure derivative $B'_{T} = 5.17$.

In contrast to the straightforward, room-temperature sound velocity measurements, high-pressure diffraction measurements are prone to error when hydrostatic conditions are not maintained. For example, the bulk modulus of the isostructural compound UMn\textsubscript{2}, reported in the same study of Itié et al.~\cite{itie1986}, was later shown by Lindbaum et al.~\cite{lindbaum2000} to be overestimated by roughly a factor of two, because line broadening masked a structural distortion and yielded an incorrect unit-cell volume. As no subsequent re-examination of high-pressure data exists for UFe\textsubscript{2}, and because two independent ultrasonic measurements agree closely at $B \approx 135$~GPa, we regard $\gamma = 1.85$ as the more reliable value.

\begin{figure}[h]
  \centering
  \includegraphics[width=0.6\linewidth]{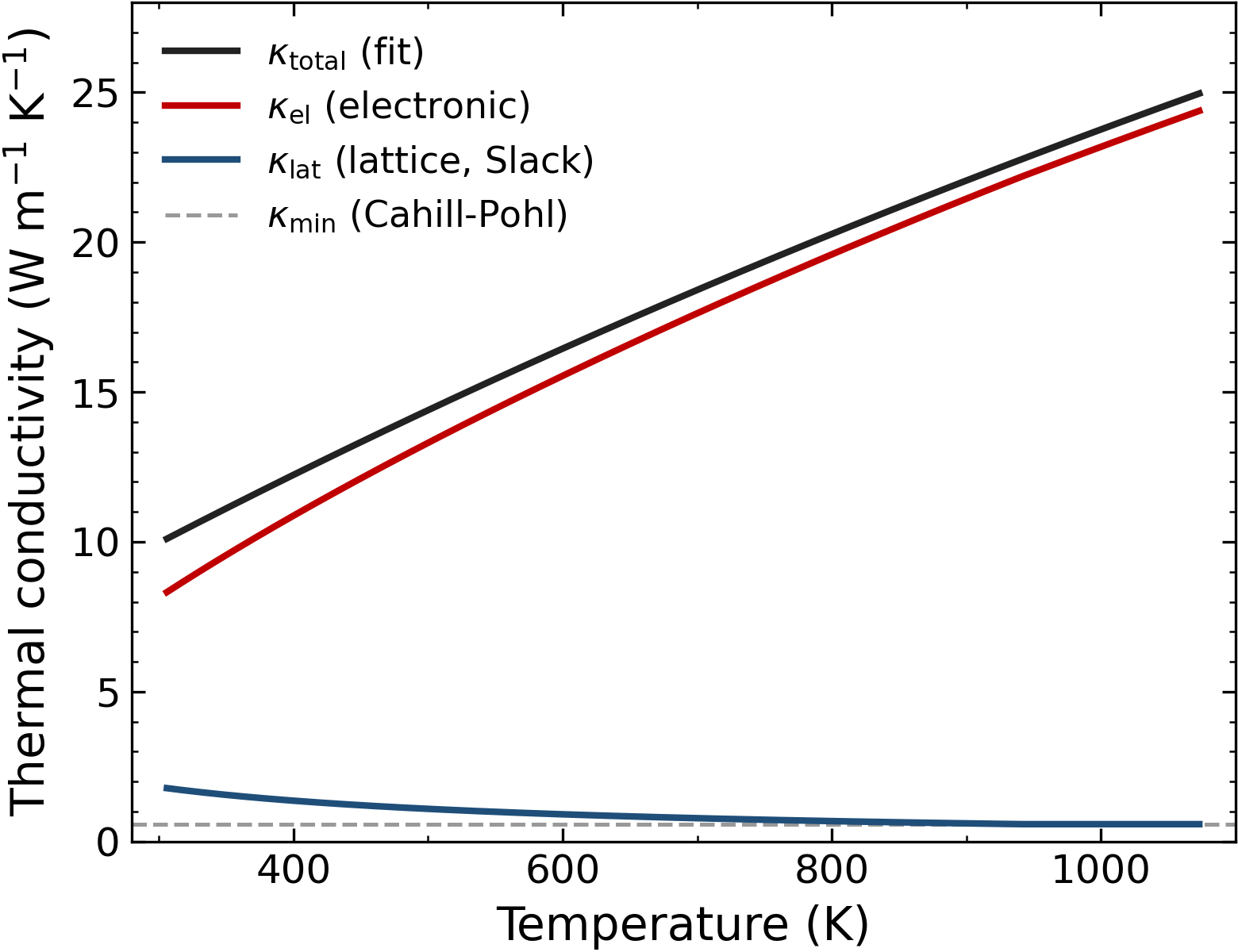}
  \caption{\textbf{Decomposition of the thermal conductivity of UFe\textsubscript{2} into electronic and lattice contributions.} The total conductivity ($\kappa_{\mathrm{total}}$, represented by its empirical fit) is separated into an electronic part ($\kappa_{\mathrm{el}}$, obtained by subtraction) and a lattice part ($\kappa_{\mathrm{lat}}$) estimated from the Slack relation~\cite{slack1979}; the Cahill--Pohl minimum ($\kappa_{\min}$)~\cite{cahill1992} is shown for reference. In the Slack estimate, the Grüneisen parameter and Debye temperature are held constant at their room-temperature values across the entire temperature range.}
  \label{fig:kappa-decomposition}
\end{figure}

Using this Grüneisen parameter together with the Debye temperature, we then partition the measured thermal conductivity into its lattice and electronic contributions (Figure~\ref{fig:kappa-decomposition}). The lattice component is estimated from the Slack relation~\cite{slack1979}, which is applicable across the measured range since $T > \theta_D$,
\begin{equation}\label{eq:slack}
\kappa_{\mathrm{lat}} = A\,\frac{\bar{M}\,\theta_{D}^{3}\,\delta}{\gamma^{2}\,n_{\mathrm{c}}^{2/3}\,T},
\end{equation}
where $\bar{M}$ is the mean atomic mass (in amu), $\theta_{D}$ the Debye temperature, $\delta$ the cube root of the mean atomic volume (in \AA), $n_{\mathrm{c}}$ the number of atoms per primitive cell, $\gamma$ the Grüneisen parameter, and $A \approx 3.1\times10^{-6}$ a numerical constant. This yields $\kappa_{\mathrm{lat}} = 1.82$~W~m$^{-1}$~K$^{-1}$ at 300~K. For comparison, the amorphous (Cahill--Pohl) minimum at $T \gg \Theta_i$~\cite{cahill1992},
\begin{equation}\label{eq:cahill}
\kappa_{\min} = \frac{1}{2}\left(\frac{\pi}{6}\right)^{1/3} k_{B}\, n_{\mathrm{a}}^{2/3}\,(v_{L} + 2v_{S}),
\end{equation}
where $\Theta_i$ is each phonon branch's Debye temperature and $n_{\mathrm{a}}$ is the atomic number density, gives $\kappa_{\min} = 0.58$~W~m$^{-1}$~K$^{-1}$. The Slack estimate lies above this lower bound, confirming that the calculated lattice conductivity is physically reasonable. Subtracting this small phonon contribution from the measured conductivity leaves an electronic part of $\approx$8.4~W~m$^{-1}$~K$^{-1}$ at room temperature, which remains the dominant contribution across the entire measured range (Figure~\ref{fig:kappa-decomposition}). This validates our choice of an electron-transport-dominated fitting equation (Equation~\ref{eq:kappa-fit}) and quantitatively confirms our earlier interpretation that the low Debye temperature ($\approx$190~K) keeps the phonon contribution small and only weakly temperature-dependent. Nevertheless, because the calculated lattice term here rests on the Slack estimate, precise electrical conductivity measurements are necessary in future work to quantify the electronic component independently through the Wiedemann--Franz law.

\subsection{Mechanical properties of fuel debris components}

From a decommissioning perspective, while the individual elastic moduli reported above are necessary for detailed thermo-mechanical simulations, they do not provide an intuitive measure of how each phase will respond to physical contact. Here, we further derive the Pugh ratio $B/G$, which essentially measures a material's resistance to volume change relative to shape change. A high $B/G$ implies ductile behavior, in which a material yields and accommodates stress before fracturing, whereas a low $B/G$ indicates brittle behavior, in which it fractures readily before significant plastic deformation. The empirical boundary between the two regimes lies near $B/G = 1.75$~\cite{pugh1954relations}. Because debris retrieval entails extensive mechanical contact such as core boring, gripping, and crushing, this distinction between ductile and brittle phases is of direct practical relevance.

\begin{figure}[h]
  \centering
  \includegraphics[width=0.6\linewidth]{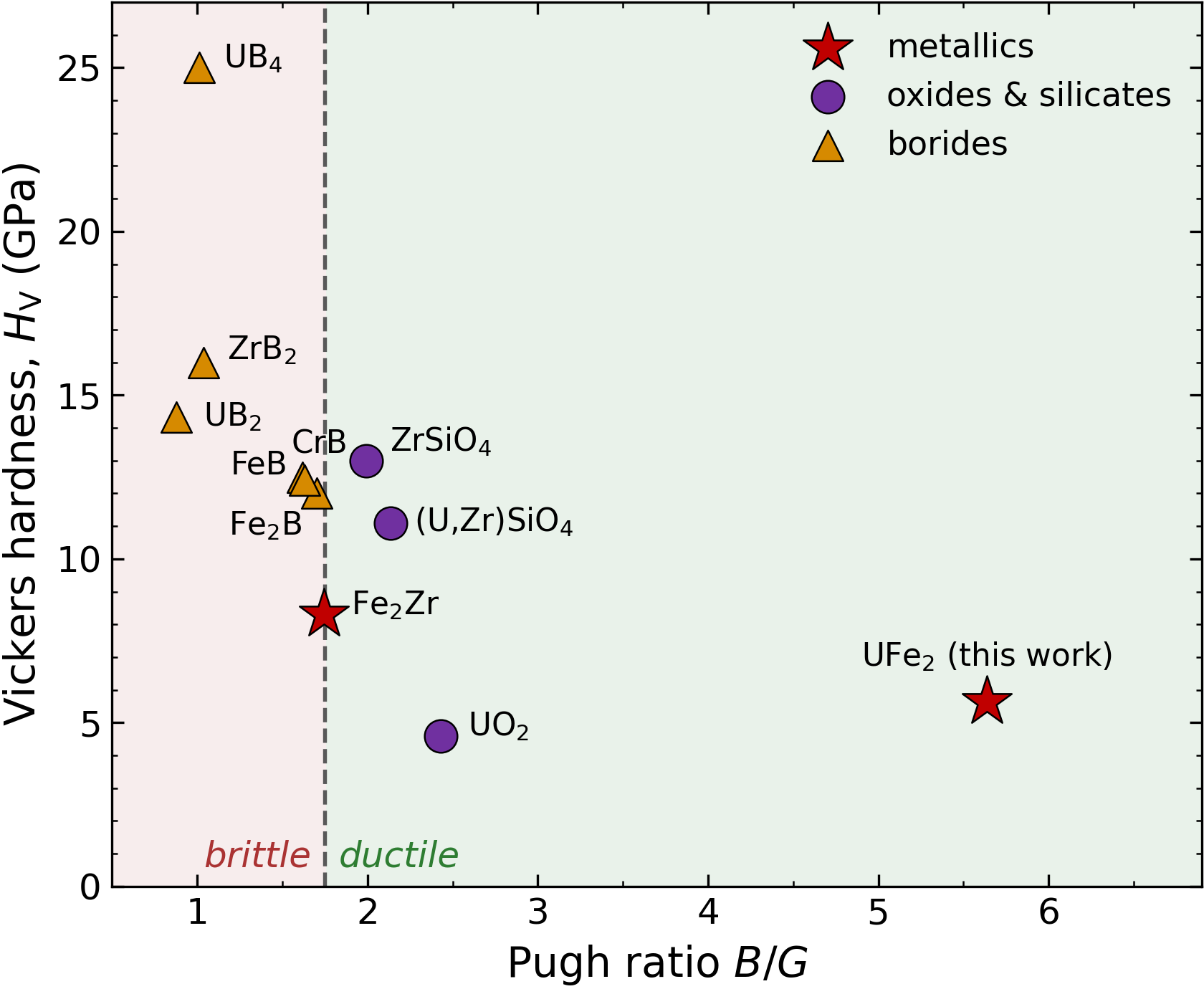}
  \caption{\textbf{Hardness versus Pugh ratio $B/G$ for UFe\textsubscript{2} and the principal phases expected in fuel debris.} The dashed line marks the ductile--brittle boundary at $B/G = 1.75$. UFe\textsubscript{2} occupies the soft, ductile region (lower right), whereas the borides are hard and brittle (upper left) and the oxide and silicate phases are intermediate.}
  \label{fig:pugh-map}
\end{figure}

We then compile the mechanical properties of these phases and map them in Figure~\ref{fig:pugh-map} (Vickers hardness versus the Pugh ratio), to compare UFe\textsubscript{2} with the other phases anticipated in fuel debris. UFe\textsubscript{2} stands far apart from every other compound, with a Pugh ratio of 5.6 and a Vickers hardness of 5.6~GPa, occupying the soft, ductile corner of the map. UO\textsubscript{2}~\cite{kato2015thermal,yamada1998mechanical} and silicates ZrSiO\textsubscript{4}~\cite{nakamori2017zrsio4} and (U,Zr)SiO\textsubscript{4}~\cite{ohishi2021mechanical}, the latter a major molten core-concrete interaction product, occupy an intermediate range with $B/G$ of approximately 2.0--2.4; borides such as Fe\textsubscript{2}B~\cite{nakamori2016mechanical}, FeB and CrB~\cite{ohishi2019thermophysical}, ZrB\textsubscript{2}~\cite{nakamori2015mechanical}, and UB\textsubscript{4} and UB\textsubscript{2}~\cite{kardoulaki2020thermophysical,matterson1961study} are the most brittle, clustering below the ductile threshold while simultaneously displaying elevated Vickers hardness. The Laves phase Fe\textsubscript{2}Zr~\cite{okada2019thermal} lies essentially at the boundary.

This contrast has several implications for debris retrieval and handling. Notably, based on the present data (Fe\textsubscript{2}Zr and UFe\textsubscript{2}), the metallic phases span a particularly wide range of mechanical behavior, so being metallic does not in itself guarantee ductile behavior. This variability highlights the importance of systematic characterization of the metallic phases. More generally, a clear ordering emerges across the debris constituents, with UFe\textsubscript{2} markedly ductile and soft, the oxide and silicate ceramics intermediate, and the borides both the hardest and the most brittle. Given such heterogeneity, characterizing the spatial distribution of these phases before retrieval would help to anticipate how the debris responds to cutting and crushing and to inform the selection of retrieval strategies. For example, although the soft, ductile UFe\textsubscript{2} offers little resistance to a cutting tool, it tends to deform rather than fracture, and so may be comparatively difficult to break into fragments by impact or crushing.

\section{Conclusion}
In this study, nearly single-phase UFe\textsubscript{2} was synthesized by arc melting followed by spark plasma sintering, and its thermal and mechanical properties were characterized from room temperature to 1073~K. Its high-temperature stability was evaluated by TG-DTA, which showed that it remains stable under an inert atmosphere over this range but oxidizes readily in air, while HT-XRD gave a linear thermal expansion coefficient of $12.5\times10^{-6}~\mathrm{K^{-1}}$ between room temperature and 473~K.

The thermal conductivity of UFe\textsubscript{2} rose monotonically from 10~W~m$^{-1}$~K$^{-1}$ near room temperature to 25~W~m$^{-1}$~K$^{-1}$ at 1073~K, dominated by its electronic component and surpassing those of Fe\textsubscript{2}B and Fe\textsubscript{2}Zr at elevated temperatures. Mechanically, UFe\textsubscript{2} is compliant, with a Young's modulus of 69~GPa and a shear modulus of 24~GPa, significantly lower than those of UO\textsubscript{2} and Fe\textsubscript{2}B. It also has a low Vickers hardness of 5.6~GPa and a high Poisson's ratio of 0.42, making it one of the most compliant and ductile phases anticipated in fuel debris, in clear contrast to the hard, brittle borides. The thermophysical data on UFe\textsubscript{2} are expected to contribute to reliable thermal and structural evaluations of fuel debris and to support the development of future retrieval technologies for the decommissioning of 1F.

\section*{Declaration of competing interest}
The authors declare that they have no known competing financial interests or personal relationships that could have appeared to influence the work reported in this paper.

\section*{Funding}
This work was supported in part by the MEXT Innovative Nuclear Research and Development Program, grant numbers 20354330 and 22682541.

\section*{Data availability}
The data that support the findings of this study are available from the corresponding author upon reasonable request.

\bibliographystyle{unsrtnat}
\bibliography{references}

\end{document}